# Experimental quantification of electronic symmetry breaking through orbital hybridization phase

Shungo Aoyagi[1,2] [†], Shunsuke Kitou[3], Yuiga Nakamura[4], Taka-hisa Arima[3,5] and Naoya Kanazawa[1] [†]

[1] *Institute of Industrial Science, University of Tokyo, Tokyo 153-8505, Japan*

[2] *Department of Applied Physics, University of Tokyo, Tokyo 113-8656, Japan*

[3] *Department of Advanced Materials Science, University of Tokyo, Kashiwa, Chiba 277-8561, Japan*

[4] *Japan Synchrotron Radiation Research Institute (JASRI), SPring-8, Hyogo 679-5198, Japan*

[5] *RIKEN Center for Emergent Matter Science, Wako 351-0198, Japan*

[†] To whom correspondence should be addressed. E-mail: s-aoyagi@iis.u-tokyo.ac.jp, naoya-k@iis.u-tokyo.ac.jp

**ABSTRACT**

Symmetry classification of crystal structures has been central to predicting physical properties of materials. While such structural classification identifies which physical responses are symmetry-allowed, the magnitudes of these responses are governed by the degree of symmetry breaking in the electronic state. However, a well-defined quantitative descriptor for the electronic symmetry breaking has been established only in limited cases such as electric polarization and magnetization. No analogous descriptor exists for most other types, including chirality. Here, we propose an experimental framework for quantifying electronic symmetry breaking from the anisotropy of valence electron density distribution. We show that the orbital hybridization phases governing this anisotropy can be uniquely determined under site symmetry constraints. Applying this framework to structurally chiral transition-metal silicides, we determine hybridization phases from their valence electron densities observed by synchrotron X-ray diffraction. From the obtained complex hybridization, we quantify an electronic chirality $\chi$ and theoretically demonstrate that it is directly proportional to circular dichroism, establishing $\chi$ as a predictive descriptor of chiral responses. This approach is systematically applicable to various point groups, offering a general route to quantifying electronic symmetry breaking and predicting associated physical properties.

**Main text**

Symmetry breaking determines the emergence of physical phenomena across a wide range of scales. Crystallographic symmetry classification according to Neumann's principle has played a central role in condensed matter physics[1], guiding the exploration of diverse physical responses as exemplified in multiferroics[2–4] and altermagnets[5,6]. Such classification, however, is inherently qualitative, just identifying allowed responses without predicting their magnitude. Designing and controlling giant responses requires a quantitative measure of the degree of symmetry breaking. For electronic properties, this demands evaluation of symmetry breaking in the electronic state itself.

As with crystallographic (structural) symmetry breaking, electronic symmetry breaking takes various forms, characterized by combinations of symmetry operations. While the degree of structural symmetry breaking can be quantified by atomic displacements[7], quantifying the degree of electronic symmetry breaking is far more challenging, because appropriate experimental descriptors are not generally obvious. Among the known forms, electric polarization serves as a descriptor for polar symmetry breaking: it measures how far the electron cloud displaced in one direction[8,9], and is directly accessible through macroscopic measurements. Similarly, magnetization quantifies time-reversal symmetry breaking, capturing the net imbalance in electron angular momentum. The availability of these descriptors has been instrumental in understanding and designing functional responses associated with polar and magnetic orders.

By contrast, for many other types of symmetry breaking, no experimentally accessible descriptor has been established, preventing quantitative evaluation of their magnitude. As theoretical alternatives, continuous symmetry measures[10,11] for molecular systems and multipoles for general electronic states[12–14] have been proposed. Although these have contributed to a coherent understanding of phenomena[15,16], they rely on knowledge of the electronic wavefunction, whose direct experimental determination remains difficult. A fundamental obstacle is that observables are expressed as expectation values of operators, from which the phase of the underlying quantum state is generally absent. Establishing an experimental method to recover phase information and quantify the degree of symmetry breaking remains an open challenge.

Here, we demonstrate that the degree of electronic symmetry breaking can be quantitatively evaluated from the anisotropy of valence electron density (VED). We adopt a density matrix representation of the electronic state, in which each off-diagonal element directly corresponds to a specific orbital hybridization term contributing to the electron density anisotropy. Under site symmetry constraints, the number of independent hybridization terms is substantially reduced, and their magnitude and phase become uniquely determinable from the observed VED. We term this approach Complex Hybrid Orbital Decomposition (CHOD) method and systematically classify the determinability of hybridization terms across all crystallographic point groups. As a concrete application, we take chirality as a representative example of symmetry breaking, motivated by the diverse chiral phenomena discovered in recent years[17–20]. Experimentally, we observe the valence electron densities of structurally chiral B20-type compounds MSi (M = Cr, Mn, Fe, Co) by synchrotron X-ray diffraction. Applying the CHOD method, we define an electronic chirality $\chi$ from the hybridization phases and analytically show its direct proportionality to circular dichroism, establishing $\chi$ as a predictive descriptor of chiral responses. This extends the role of X-ray diffraction from structural characterization to quantitative evaluation of electronic symmetry breaking and prediction of associated physical properties.

**Complex hybrid orbital decomposition method**

Electron density, unlike macroscopic observables such as electric polarization or magnetization, is spatially resolved and carries far richer information about the underlying quantum state. We focus on the VED around atomic nuclei, where atomic orbitals $\psi_i(r)$ serve as a natural basis set. The spatial anisotropy of the VED is not merely a qualitative feature of orbital occupation but quantitatively encodes the orbital hybridization, including the phase difference between contributing orbitals. This can be seen most simply in a single-electron state $\psi(r) = c_1\psi_1(r) + c_2\psi_2(r)$. The electron density $|\psi(r)|^2$ contains the interference term $2|c_1||c_2|\mathrm{Re}[e^{i(\phi_2-\phi_1)}\psi_1^*\psi_2]$, where $\phi_1$ and $\phi_2$ are the phases of $c_1$ and $c_2$. When $\psi_1$ and $\psi_2$ are eigenstates of orbital angular momentum, the phase difference $\Delta\phi = \phi_2 - \phi_1$ manifests itself as a rotation of VED (Fig. 1**a**). Consequently, both the amplitudes and the phase difference are in principle extractable from the shape of the observed electron density. This is the central principle of our approach.

Despite its conceptual simplicity, such extraction has not been explored beyond a few organic molecules[21]. For general systems including inorganic crystals, the involvement of d and f orbitals greatly increases the number of relevant orbitals and hence interference terms, whose superposition produces an intricate anisotropic angular pattern of the VED. While such an anisotropic pattern still encodes the hybridization amplitudes and phases, their extraction from an N-particle wavefunction is intractable in practice. We address this difficulty by adopting the one-electron reduced density matrix (1-RDM)[22] for valence electrons. The 1-RDM fully describes one-body quantities such as electron density regardless of the complexity of the underlying many-electron state, while also incorporating thermal and statistical mixing effects.

The 1-RDM $P(\mathbf{r},\mathbf{r}')$ can be approximated by the following quadratic form[23,24]:

$$P(\mathbf{r},\mathbf{r}') = \sum_{ij} P_{ij}\psi_i^*(\mathbf{r})\psi_j(\mathbf{r}'), \qquad (1)$$

where $P_{ij} = P_{ji}^*$. Setting $\mathbf{r} = \mathbf{r}'$ yields the electron density $\rho(\mathbf{r}) = P(\mathbf{r},\mathbf{r})$, in which the off-diagonal elements $P_{ij}\psi_i^*(\mathbf{r})\psi_j(\mathbf{r})$ ($i \neq j$) correspond to orbital hybridization terms, generalizing the interference term of the single-electron case. The CHOD method decomposes the anisotropic angular pattern of the observed VED into these hybridization terms with complex-valued coefficients (Fig.

**1b**). Atomic orbitals as basis functions are first categorized by irreducible representations of the site symmetry to reduce the number of independent terms. The coefficient matrix $P$ is optimized by fitting the calculated $\rho(\mathbf{r})$ to the observation. (See Supplementary Note 1 for details of the CHOD method.)

To obtain the VED from X-ray diffraction data, we used the core-differential Fourier synthesis (CDFS) method[25,26]. The contribution of core electrons with spherically symmetric distribution is calculated and subtracted from the measured structure factors, and the remaining VED is reconstructed through the Fourier synthesis (see Methods and Supplementary Note 1 for details).

The determinability of hybridization phases depends on the site symmetry. To illustrate how symmetry constraints make the phase extraction feasible, consider a site with $C_3$ symmetry. The irreducible representations of the hybridization terms are $A$, $E_1$, and $E_2$, or generally, $A$ and $E$ (Fig. 2**a**). Here, $E_1$ and $E_2$ are a complex conjugate pair; they are equivalent as long as time-reversal symmetry is considered. Hereafter, we consider only $A$ and $E_1$ when $s$, $p$, and $d$ orbitals are used as basis functions. The $A$ representation preserves continuous rotational symmetry about the principal axis and therefore does not contribute to the three-fold anisotropy. Within the irreducible representation $E_1$, the hybridization terms reduce to three types (Fig. 2**b**). Among them, the terms involving orbitals with different magnetic quantum numbers $m$, specifically $P_{12}^{(E)}(p_{m=1}^* \times d_{m=-2})$ and $P_{32}^{(E)}(d_{m=1}^* \times d_{m=-2})$, contribute to the three-fold rotational anisotropy. Here, the basis functions $\psi_i^{(E)}(\mathbf{r})$ ($i$ = 1, 2, and 3) belong to the irreducible representation $E$ and correspond to $p_{m=1}$, $d_{m=-2}$, and $d_{m=1}$, respectively. Similarly, $\psi_i^{(A)}(\mathbf{r})$ ($i$ = 1, 2, and 3) belong to the irreducible representation $A$ and correspond to $s$, $p_z$, and $d_{3z^2-r^2}$, respectively. When these two terms are summed with real coefficients, the resulting electron density retains three-fold rotational symmetry but is aligned straight along the principal axis, yielding a non-chiral distribution (Fig. 2**c**). When a phase factor is introduced to one term, the hybridization term rotates about the principal axis, twisting the three-fold pattern into a chiral distribution (left panel in Fig. 2**d**). Applying the opposite-signed phase factor reverses the handedness of the VED twist (right panel in Fig. 2**d**). In this manner, the phase of each hybridization term controls its azimuthal orientation, and the phase difference between the two hybridization terms governs the degree of chiral twisting of the electron density. Conversely, the rich spatial information encoded in such anisotropic distributions allows the hybridization phases to be uniquely extracted from the experimentally observed electron density. We have systematically classified the determinability of hybridization terms for all crystallographic point groups in Supplementary Note 2.

**Orbital decomposition of chiral electron density**

As a concrete application of the CHOD method, we analyze the VED distribution of an enantiomer (B-form) CoSi[27] (Fig. 3**a, b**). CoSi crystallizes in the $B$20-type structure with space group $P2_13$. In this chiral cubic structure, both Co and Si occupy crystallographic sites with $C_3$ point symmetry. Here, we define the A-form of *B*20-type crystals as the structure in which transition metal sites are stacked clockwise along the [111] direction, and the B-form as its mirror image. CoSi hosts multifold topological fermions near the Fermi level and serves as a representative material for studying chiral electronic responses[28–34].

The VED around the Co site on the [111] axis exhibits two characteristic types of asymmetry: a chiral shape, twisted with respect to the three-fold axis (Fig. 3**c**, viewed from the $[\bar{1}\bar{1}\bar{1}]$ direction); and polar asymmetry, with the distribution biased along the [111] direction (Fig. 3**d**, side view). This polar component requires the inclusion of odd-parity 4*p* orbitals in the basis set alongside even-parity 3*d* and 4*s* orbitals. We applied the CHOD method with this basis. The reconstructed VED (Fig. 3**e, f**) reproduces both the chiral twist and the polar asymmetry. The weighted relative error, defined as the square root of the objective function, is 1.3% (see Supplementary Note 3 for details).

To examine how each hybridization term contributes to the VED anisotropy, we compare the observed and reconstructed VED along three line cuts indicated in Fig. 3**b**. Line 1 is a radial cut along the principle axis, showing different peak heights on opposite sides of the Co site (Fig. 3**g**), indicating polar asymmetry. The decomposition (Fig. 3**h**) reveals that this radial asymmetry originates exclusively from hybridization terms between orbitals of opposite parity (also see Fig. 2**b**), specifically $P_{12}^{(A)}(s^* \times p_z)$ (purple curve in Fig. 3**h**) and $P_{32}^{(A)}(d_{z^2}^* \times p_z)$ (green curve in Fig. 3**h**). Lines 2 and 3 are azimuthal cuts at equal distances (0.241 Å) from the Co site and constant polar angles ($\theta = \frac{\pi}{4}$ and $\frac{3\pi}{4}$), respectively. Both profiles exhibit sinusoidal three-fold modulations (Fig. 3**i, k**), governed by the hybridization terms $P_{12}^{(E)}(p_{m=1}^* \times d_{m=-2})$ (red curve in Fig. 3**j, l**) and $P_{32}^{(E)}(d_{m=1}^* \times d_{m=-2})$ (blue curve in Fig. 3**j, l**). The decomposition shows that $P_{12}$ and $P_{32}$ contribute with a mutual phase difference in both Line 2 and Line 3 (Fig. 3**j, l**). Moreover, this phase difference takes distinct values for Lines 2 and 3, because $P_{32}^{(E)}(d_{m=1}^* \times d_{m=-2})$ reverses sign between the $\theta < \frac{\pi}{2}$ and $\theta > \frac{\pi}{2}$ sides of the Co site whereas $P_{12}^{(E)}(p_{m=1}^* \times d_{m=-2})$ does not (also see Fig. 2**b**). The resulting opposite azimuthal rotations on the two sides give rise to the overall chiral twist along the principal axis, consistent with the model case of Fig. 2**d** (see Supplementary Note 4 for the mathematical details).

We validate the analysis through several consistency checks. Excluding 4*p* orbitals from the basis restricts the VED to an inversion-symmetric distribution, inconsistent with the observation (Extended Data Fig. 1). Restricting the coefficients $P_{ij}$ to real values reproduces the polar asymmetry but fails to express the chiral twist (Extended Data Fig. 2). These control analyses confirm that both 4*p* orbital inclusion and complex-valued coefficients are essential for describing the VED of CoSi.

**Quantification of electronic chirality**

We applied the CHOD method to the VED of four *B*20-type *M*Si (*M* = Cr, Mn, Fe, Co) (Extended Data Fig. 3; see Supplementary Notes 3 and 5 for details of the analyses), including FeSi and CoSi reported in our previous study[27]. These compounds are isostructural with only slight differences in the lattice parameter and atomic coordinates. The rigid-band approximation holds well across the series[35], allowing systematic comparison of electronic chirality as a function of electron filling while the structural chirality remains essentially fixed[36].

Figure 4**a**, **b** shows the magnitude and phase of the two hybridization coefficients $P_{12}^{(E)}$ and $P_{32}^{(E)}$, which govern the chiral twist of the VED as discussed in Fig. 3**i**–**l**. The reconstructed VED (Fig. 4**c**) faithfully reproduces the observed VED for all four compounds (Extended Data Fig. 3), and reveals that the degree of chiral twisting varies markedly across the series. To understand this variation in terms of the hybridization coefficients, we compare specific pairs of compounds. MnSi and CoSi exhibit comparable phase differences between $P_{12}^{(E)}$ and $P_{32}^{(E)}$, but differ markedly in the product of their amplitudes, resulting in pronounced differences in the magnitude of the chiral twist. In contrast, FeSi and CoSi have comparable amplitude products, yet their phase differences differ substantially: approximately $\pi$ for FeSi and $\pi/2$ for CoSi. A phase difference of $\pi$ yields real-valued coefficients that produce an achiral distribution, whereas a phase difference of $\pi/2$ introduces a maximal imaginary component, producing the strongest chiral twist. These comparisons indicate that the degree of electronic chirality is determined by both the product of amplitudes and the phase difference.

Here we note that the amplitude variations across the series can be understood from electron filling. $P_{12}$ involves hybridization with $4p$ orbitals, which lie at higher energy than $3d$ orbitals. Since CoSi has the highest $3d$ electron count in the series, more electrons are available to occupy the $4p$ orbitals, making the contribution of $P_{12}^{(E)}$ largest in CoSi. $P_{32}^{(E)}$ represents $d$–$d$ hybridization that vanishes when the $d$ shell is either empty or fully occupied, and reaches a maximum at intermediate filling, which corresponds to FeSi in this series.

Accordingly, we define a descriptor of the electronic chirality $\chi$ for systems with $C_3$ symmetry composed of $d$ and $p$ orbitals as follows:

$$\chi = \left|P_{32}^{(E)}\right|\left|P_{12}^{(E)}\right|\sin(\arg(P_{32}^{(E)}) - \arg(P_{12}^{(E)})).$$

Figure 4**d** shows $\chi$ for the four A-form *B*20-type compounds, mapped as a function of the amplitude product and the phase difference. Although the sign of $\chi$ does not change among compounds with identical structural chirality, the magnitude varies considerably, with CoSi exhibiting the largest $\chi$. Furthermore, in B-form CoSi, the sign of $\chi$ is reversed corresponding to the inversion of crystallographic chirality, while its magnitude remains unchanged. This quantitative measure confirms the systematic variation of chiral twisting observed across the series (Fig. 4**c**), demonstrating that $\chi$ serves as a well-defined descriptor of electronic chirality.

The electronic chirality $\chi$ is not only a geometric measure of electron density but also a quantitative indicator of chiral physical properties. Since the CHOD analysis determines the occupied orbital components from the VED, the unoccupied components are obtained by subtraction from the full atomic orbital set. This enables estimation of the circular dichroism (CD) for optical transitions in

chiral systems. The frequency-integrated differential absorption coefficient can be derived as:

$$\int \Delta\alpha(\omega)\, d\omega \propto \chi,$$

where $\Delta\alpha(\omega) = \alpha_L(\omega) - \alpha_R(\omega)$ is the difference in absorption coefficients for left and right circularly polarized light (see Supplementary Note 6 for derivation). The integration is performed over the entire frequency range spanned by transitions between occupied and unoccupied states within the $3d$ and $4p$ manifold. This proportionality demonstrates that $\chi$ can predict the magnitude of chiral optical responses, establishing $\chi$ as an experimentally accessible descriptor of electronic chirality, analogous to electric polarization and magnetization. While the frequency-integrated CD in transmission geometry is inaccessible for metallic materials, $\chi$ is measurable for any crystalline material. Similar descriptors can be defined for various chiral crystals, though modifications may be necessary depending on the space group and the involvement of $f$ orbitals.

**Summary and perspectives**

We have proposed a general experimental framework for quantitatively evaluating electronic symmetry breaking from the anisotropy of valence electron density. The CHOD method decomposes the observed VED into orbital hybridization terms with complex-valued coefficients, enabling direct determination of both the magnitude and phase of hybridization from diffraction data. Applying this framework to $B$20-type chiral monosilicides, we defined the electronic chirality $\chi$ and theoretically demonstrated its direct proportionality to circular dichroism, establishing $\chi$ as an experimentally measurable descriptor of electronic chirality analogous to electric polarization and magnetization.

The applicability of this framework extends beyond chirality. The orbital hybridization-based approach can be generalized to arbitrary symmetry breaking described by higher-order electrical multipoles[12]. Furthermore, by applying analogous analysis to spin-resolved electron density obtained through techniques such as neutron diffraction[37], symmetry breaking associated with magnetic order, such as in altermagnets[5,6] and piezomagnets[38], could also be quantitatively described. This generalization would provide experimental access to descriptors for hidden symmetry breaking that have so far been evaluated only through theoretical calculations. Combined with measurements in various circumstances including external fields, the CHOD method will provide a foundation for designing and controlling giant physical property responses based on the quantified magnitude of electronic asymmetry.

**Methods**

**Sample preparation**

The single crystals of *B*20 silicides were grown by the chemical vapor transport (CVT) method[27]. Stoichiometric amounts of the transition metal and silicon elements were arc-melted to achieve a homogeneous mixture. The resulting melt was then ground into a powder form, which was loaded into a quartz tube along with a transport agent ($I_2$) and subsequently sealed under vacuum. For CrSi, a temperature gradient was applied with the high-temperature zone at 1100 °C and the low-temperature zone at 1000 °C, whereas for MnSi, the high- and low-temperature zones were set at 1000 °C and 900 °C, respectively. See Ref. 26 for the growth temperature conditions for FeSi and CoSi. The samples were grown under these conditions for three weeks, typically yielding single crystals on the order of 100 μm in size. The crystalline chirality is defined such that, in the A-form (B-form) crystal, the transition metal atoms are stacked in a clockwise (counterclockwise) arrangement along the [111] direction.

**XRD experiments**

The XRD experiments were performed on BL02B1 at a synchrotron facility SPring-8[39], Japan. An $N_2$-gas-blowing device was used to cool the crystal to 100 K. A two-dimensional detector CdTe PILATUS, which had a dynamic range of $\sim 10^6$, was used to record the diffraction patterns. The intensities of Bragg reflections of the interplane distance $d > 0.28$ Å were collected by the CrysAlisPro program[40] using a fine slice method, in which the data were obtained by dividing the reciprocal lattice space region in increments of $\omega = 0.01°$. Intensities of equivalent reflections were averaged, and the structural parameters were refined by JANA2006[41]. High-angle reflections ($\sin \theta/\lambda > 0.5$ Å$^{-1}$) were exclusively used for structural refinement to perform high-angle analysis. Since the contribution of spatially spread valence electrons to X-ray diffraction is negligibly small in the high-angle region, the structural parameters, including the atomic displacement parameters (mainly due to the thermal vibration), are obtained with high accuracy. The crystal structures are drawn by using VESTA[42].

**CDFS analysis**

The CDFS method was used to extract the VED distribution around each atomic site[25]. Core electron configurations were assumed as [Ne] for Si, and [Ar] for both Cr and Mn atoms. The contribution of the thermal vibration was subtracted from the VED using the atomic displacement parameters determined by the high-angle analysis. In CrSi (MnSi), the voxel of the three-dimensional electron distribution is defined as a cube with a side length of 0.0231 Å (0.0227 Å). It should be noted that the absolute value of the obtained electron distribution does not directly reproduce the number of valence electrons around the atoms, partly because the double scattering, absorption, extinction, and detector saturation[43] could not be completely excluded in the measurement of diffraction intensities. We also verify the reliability of the structure factor phases used in the structural analysis and the CDFS method (see Supplementary Note 7 for details).

**CHOD analysis**

The CHOD method was applied to the valence electron density distributions obtained by the CDFS method. As basis functions, $3d$, $4s$, and $4p$ atomic orbitals were employed (for details of the orbitals, see Supplementary Note 1). The axis orientation was defined with the local three-fold axis as the principal axis, and optimization was performed within a radius of 0.6 Å from the nucleus. For basis function truncation, reflections with $d$-values smaller than the experimentally obtained resolution $d_{min}$ were set to zero. The Python library cvxpy was used as the solver for the optimization problem, with SCS as the backend.

**Acknowledgements** We thank C. Koyama for experimental support. This work was supported by JSPS KAKENHI (Grants No. 23H04017, 23H05431, 23H05462, 24H00417, 24H01644, 24H01652, 25H02126), JST FOREST (Grants No. JPMJFR2038, No. JPMJFR2362), JST CREST (Grant No. JPMJCR23O3), the Mitsubishi Foundation, the Sumitomo Foundation and FoPM (WINGS Program, the University of Tokyo). The synchrotron radiation experiments were performed at SPring-8 with the approval of the Japan Synchrotron Radiation Research Institute (JASRI) (Proposals No. 2024A1709, 2024B1599, 2024B2017, and 2025A1998).

**Author contributions** S.A. and N.K. conceived the project. S.A. and N.K. grew the single crystals. S.A. and S.K. performed XRD measurements with support from Y.N. and N.K.; S.A. analyzed the data and implement the program with support from S.K., T.A., and N.K.; N.K. organized the project. S.A. and N.K. wrote the draft and all the authors discussed the results and commented on the manuscript.

**Competing interests** The authors declare no competing interests.

**Data availability** All the data presented in figures are available at UTokyo Repository: https://...

**Figures**

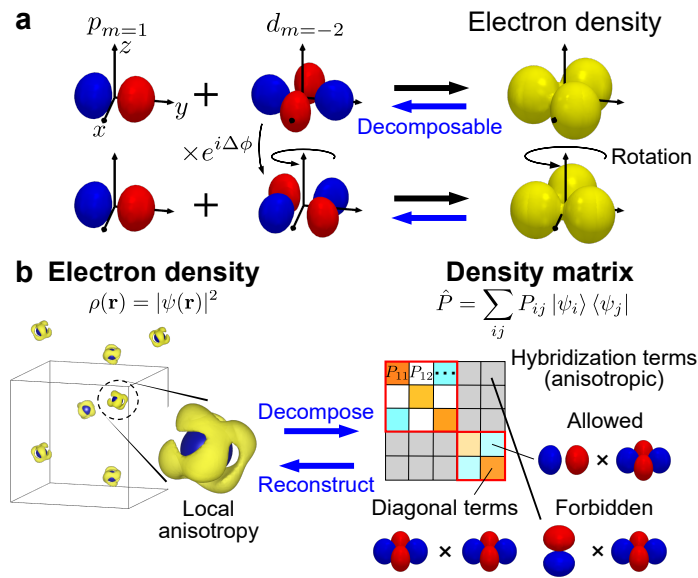

**Figure 1 | Principle of complex hybrid orbital decomposition of valence electron density distribution. a,** Electron density resulting from the superposition of $p_{m=1}$ and $d_{m=-2}$ wavefunctions with different relative phases. Applying a phase factor $e^{i\Delta\phi}$ to the $d_{m=-2}$ orbital induces a spatial rotation of the resulting electron density; conversely, this angular dependence allows the phase to be restored from the observed electron density. The red and blue lobes represent the positive and negative phases of the real part of the spherical harmonics. Black arrows indicate the $x$, $y$, and $z$ coordinate axes, with the $z$ axis chosen as the quantization axis. **b,** Schematic of the complex hybrid orbital decomposition (CHOD) method linking the observed electron density $\rho(\mathbf{r})$ to the density matrix $P$. Our method exploits symmetry constraints to limit the active components to diagonal and allowed off-diagonal hybridization terms, effectively reducing the parameter space. This approach allows us to quantitatively extract the amplitude and phase of the hybridization terms causing the electron density anisotropy. The extracted values can subsequently be used to directly reconstruct the localized electron density.

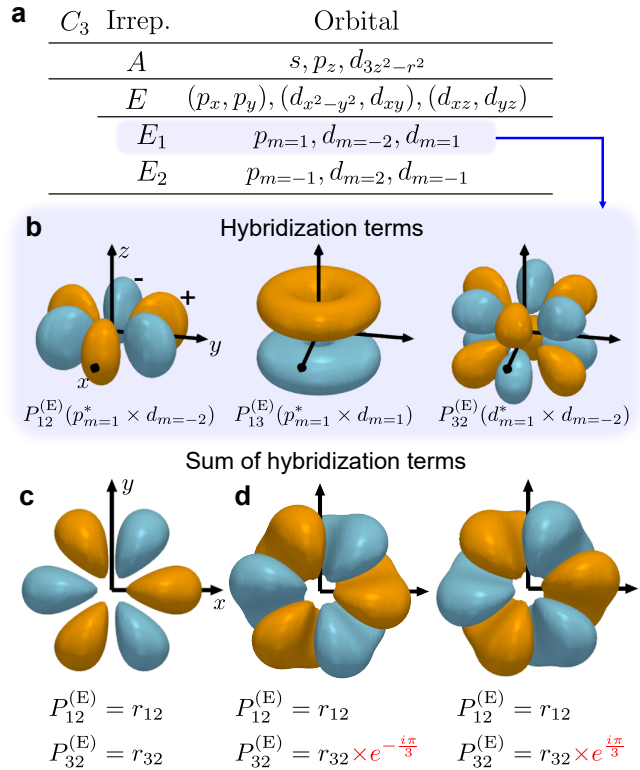

**Figure 2 | Site-symmetry constraints on orbital hybridization and the impact of phase on electron density. a**, Orbital classification under $C_3$ point group symmetry (excerpt from Supplementary Table 1). **b**, Spatial distribution of the hybridization terms (basis function products) belonging to the $E_1$ irreducible representation. Black arrows indicate the $x, y,$ and $z$ coordinate axes, with the $z$ axis chosen as the quantization axis. The orange and cyan lobes represent the positive and negative value of the hybridization terms. **c, d**, Impact of complex phases on chirality. Achiral electron density distribution resulting from real-valued parameters ($P_{12}^{(E)} = r_{12}$ and $P_{32}^{(E)} = r_{32}$) in **c**. Chiral electron density distributions emerging from complex-valued parameters with phase shifts to $P_{32}^{(E)} = r_{32} \times e^{-i\pi/3}$ (left) and $r_{32} \times e^{i\pi/3}$ (right) in **d**.

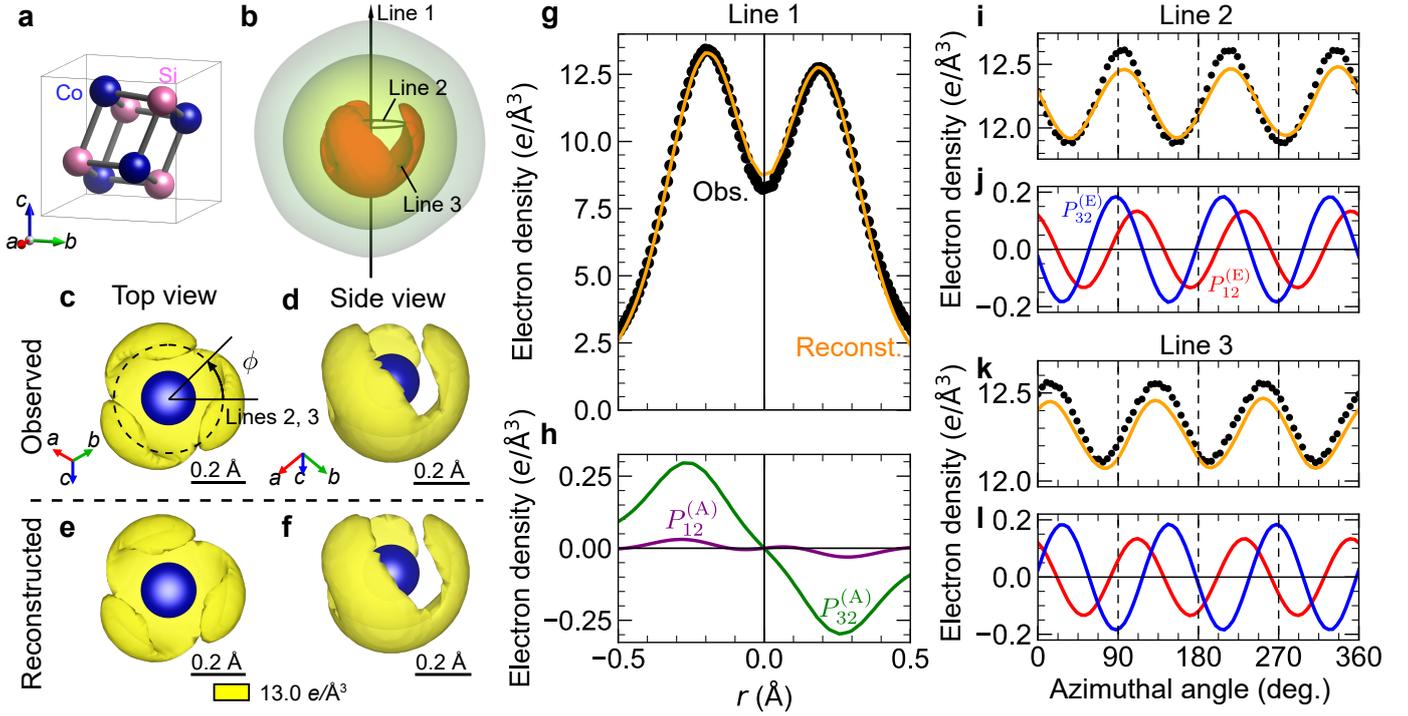

**Figure 3 | Observed chiral electron density and extracted hybridization phases in CoSi. a,** Crystal structure of B-form CoSi. **b,** Observed valence electron density (VED) distribution around a Co site of B-form CoSi. Three isosurfaces are shown at electron densities of 3.5, 7.5 and 13.0 $e$ Å$^{-3}$, from outermost to innermost. Line 1 is along the $[\bar{1}\bar{1}\bar{1}]$ direction; lines 2 and 3 are azimuthal circles of radius 0.2 Å, centered at positions displaced by +0.135 Å and −0.135 Å from the Co site along the [111] direction, respectively. **c–f,** Comparison of experimentally observed (**c, d**) and reconstructed (**e, f**) VED distributions around the Co site. The reconstruction employs a (3$d$, 4$s$, 4$p$) orbital basis. The isosurface level is set to 13.0 $e$ Å$^{-3}$ for all plots (**c–f**). Dashed circle represents the lines 2 and 3 in panel **b**. **g,** Radial distribution of the observed (black) and reconstructed (orange) VED near the Co site along line 1 in **b**. **h,** Hybridization terms ($P_{12}^{(A)}(s^* \times p_z)$; purple and $P_{32}^{(A)}(d_{z^2}^* \times p_z)$; green) contributing to the polar anisotropy seen in **g**. **i–l,** Azimuthal dependence along line 2 (**i, j**) and line 3 (**k, l**) in **b**. **i, k,** Observed (black) and reconstructed (orange) VED. **j, l,** Azimuthal dependence of the two hybridization terms ($P_{12}^{(E)}(p_{m=1}^* \times d_{m=-2})$; red and $P_{32}^{(E)}(d_{m=1}^* \times d_{m=-2})$; blue) contributing to the chiral anisotropy. The $P_{12}^{(E)}(p_{m=1}^* \times d_{m=-2})$ term remains unchanged between **j** and **l**, whereas the $P_{32}^{(E)}(d_{m=1}^* \times d_{m=-2})$ term reverses sign.

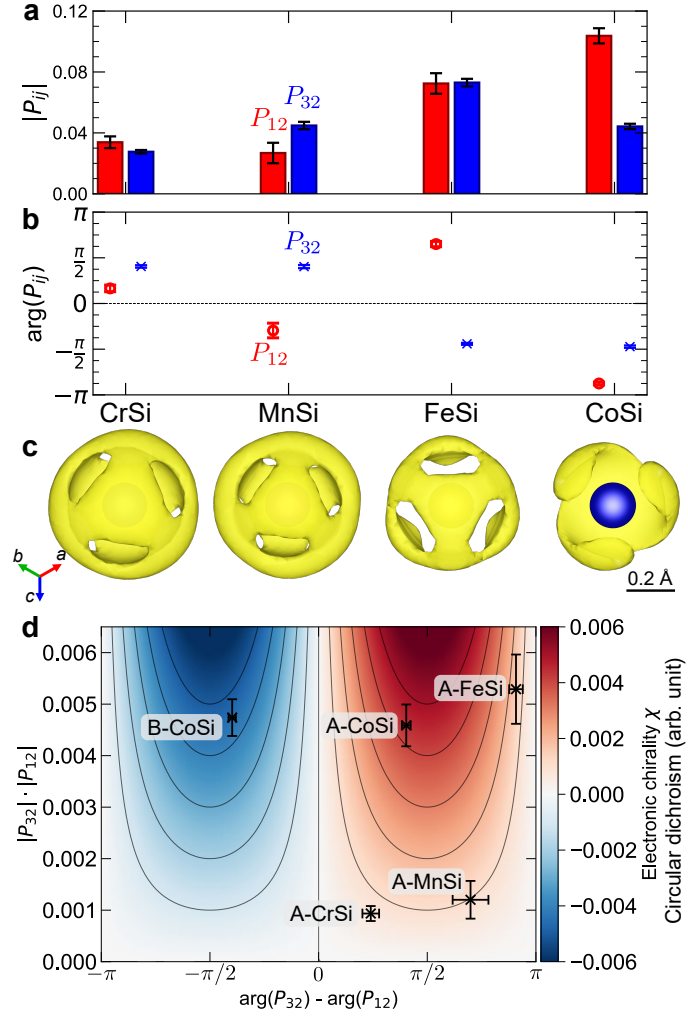

**Figure 4 | Orbital hybridization and electronic chirality in transition-metal monosilicides. a**, **b**, Evolution of the amplitude (**a**) and phase (**b**) of the density matrix components across the *B*20 transition-metal monosilicides (A-form *M*Si; *M* = Cr, Mn, Fe, Co) . Red and blue plots denote the $P_{12}^{(E)}$ and $P_{32}^{(E)}$ hybridization terms, respectively. **c**, Reconstructed valence electron density distributions for the *M*Si series viewed from $[\bar{1}\bar{1}\bar{1}]$. The isosurface levels were set to 4.81 (CrSi), 6.12 (MnSi), 8.43 (FeSi) and 10.95 (CoSi) $e$ Å$^{-3}$, respectively. **d**, Electronic chirality χ and circular dichroism mapped as functions of the phase difference (horizontal axis) and the amplitude product (vertical axis), with the experimentally determined values for A-form *M*Si (*M* = Cr, Mn, Fe, Co) and B-form CoSi overlaid. Comparison of A-form and B-form CoSi demonstrates the sign reversal of χ upon inversion of crystal chirality.

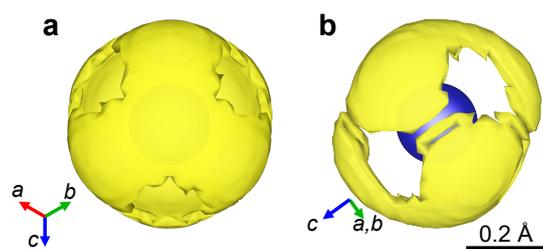

**Extended Data Fig. 1 | Reconstructed valence electron density of CoSi using the CHOD method without 4*p* orbitals. a,** Valence electron density distribution viewed from [111] direction. **b,** Valence electron density distribution viewed from [$\bar{1}$10] direction. The isosurface level is set to 13.0 $e$ Å$^{-3}$, and the scale bar represents 0.2 Å.

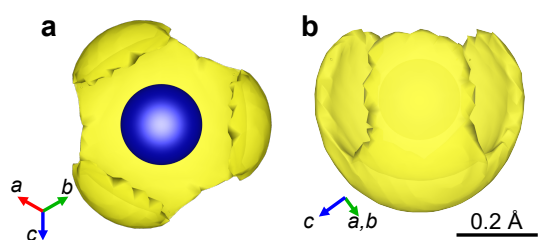

**Extended Data Fig. 2 | Reconstructed valence electron density of CoSi using the CHOD method with real-valued coefficients. a,** Valence electron density distribution viewed from [111] direction. **b,** Valence electron density distribution viewed from [$\bar{1}$10] direction. The isosurface level is set to 13.0 $e$ Å$^{-3}$, and the scale bar represents 0.2 Å.

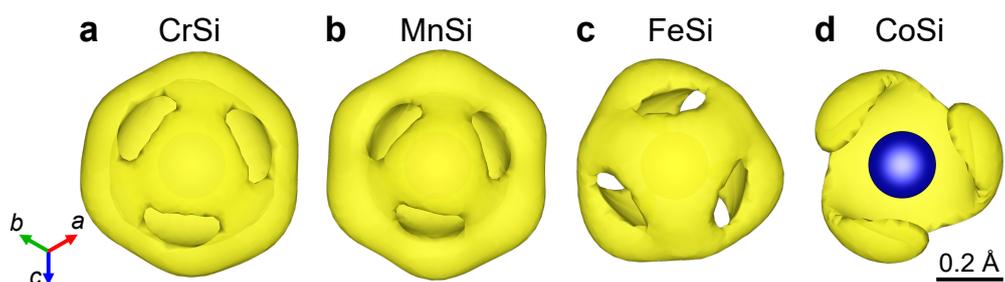

**Extended Data Fig. 3 | Observed valence electron density of A-form *B*20 compounds. a-d,** Valence electron density distribution for *M*Si (*M* = Cr, Mn, Fe, Co) viewed from [$\bar{1}\bar{1}\bar{1}$] direction. The isosurface levels were set to 4.81 (CrSi, panel **a**), 6.0 (MnSi, panel **b**), 8.43 (FeSi, panel **c**) and 11.0 (CoSi, panel **d**) $e$ Å$^{-3}$, respectively. The scale bar represents 0.2 Å.